\documentclass[%
reprint, superscriptaddress,amsmath,amssymb,aps,
prl,floatfix,showpacs]{revtex4-1}

\usepackage[FIGTOPCAP,tight]{subfigure}
\usepackage{graphicx}
\usepackage{dcolumn}
\usepackage{bm}
\usepackage{amsmath}

\usepackage{xcolor}
\newcommand{\bra}[1]{\left< #1 \right\vert}
\newcommand{\ket}[1]{\left\vert #1 \right>}
\newcommand{\pare}[1]{\left( #1 \right)}
\newcommand{\abs}[1]{\left\vert #1 \right\vert}
\newcommand{\cor}[1]{\left[ #1 \right]}

\newcommand{\llav}[1]{\left\lbrace #1 \right\rbrace}

\begin{document}

\preprint{APS/123-QED}

\title{Genuinely quantum effects in nonlinear spectroscopy: vacuum fluctuations and their induced superradiance}

\author{Roberto de J. Le\'{o}n-Montiel}
\affiliation{Department of Chemistry $\&$  Biochemistry, University of California San Diego, La Jolla, California 92093, USA}
\author{Zixuan Hu}
\affiliation{Department of Chemistry $\&$  Biochemistry, University of California San Diego, La Jolla, California 92093, USA}
\author{Joel Yuen-Zhou}\email{joelyuen@ucsd.edu}
\affiliation{Department of Chemistry $\&$  Biochemistry, University of California San Diego, La Jolla, California 92093, USA}


\begin{abstract}

The classical or quantum nature of optical spectroscopy signals is a topic that has attracted great attention recently. Spectroscopic techniques have been classified as quantum or classical depending on the light-source used in their implementations. In this way, experiments performed with quantum light---such as entangled photon pairs---have been labeled as quantum spectroscopies, whereas those performed with coherent laser pulses are generally referred to as classical ones. In this work, we highlight the fact that typical laser-spectroscopy signals should sometimes be deemed quantum too, as they contain information about the quantum vacua of the modes that interact with the sample. Using a minimalistic model, namely frequency-integrated pump-probe spectroscopy, we demonstrate that vacuum contributions can be expressed as a correction term to the \emph{classical} pump-probe signals, which scales linearly with the intensity of the pump field. Remarkably, we show that these vacuum contributions may not be negligible and lead to the observation of superradiance in pump-probe experiments, provided that fields interacting with the medium are arranged in a collinear configuration.



\end{abstract}
\maketitle


\emph{Introduction.---}Multidimensional spectroscopies have long been used in analytical and physical chemistry for extracting information about the dynamics and structure of complex systems, from small organic molecules to photosynthetic complexes and large proteins, using tools ranging from nuclear magnetic resonance to ultraviolet-visible light \cite{ernst1987,mukamel_book,hamm2011,cho2009,joel_book}. Although in the optical regime these techniques have typically been implemented with laser-light pulses, recent works suggest that the use of quantum light, such as entangled photon pairs, may open new and exciting avenues in experimental spectroscopy \cite{javanainen1990,lee2006,muthukrishnan2004,fei1997,guzman2010,saleh1998,kojima2004,roberto2013,roslyak2009,roslyak2009-2,raymer2013,schlawin2013,schlawin2016,dorfman2016,shapiro2011}. Along these lines, quantum light has enabled the prediction and observation of fascinating two-photon absorption (TPA) scenarios, such as the linear dependence of the TPA rate on photon flux \cite{javanainen1990,lee2006}, as well as the prediction of phenomena as varied as inducing disallowed spectroscopic transitions \cite{muthukrishnan2004}, two-photon-induced transparency \cite{fei1997,guzman2010}, virtual-state spectroscopy \cite{saleh1998,kojima2004,roberto2013}, control of nonlinear pulsed spectroscopy \cite{roslyak2009,roslyak2009-2,raymer2013,schlawin2016,dorfman2016,schlawin2013}, and the control of entanglement in matter \cite{shapiro2011,shapiro_book}.

In general, spectroscopy techniques have been classified as quantum or classical depending on the light-source used in their implementation \cite{mukamel2015}. Interestingly, this classification has been reinterpreted in light of recent works where quantum spectroscopy signals are obtained by properly averaging their ``classical'' counterparts \cite{kira2011,mootz2014}. Indeed, these results have suggested that there exists a fundamental connection between quantum and classical spectroscopy signals, which resides in the fact that coherent states, which are the most classical states of light, are ultimately also quantum mechanical in nature.

Motivated by those discussions \cite{mukamel2015,kira2015,mukamel2015-2}, we here present a thorough study of a minimalistic model, namely frequency-integrated pump-probe spectroscopy. We show that a consistent understanding of nonlinear spectroscopy with ``classical'' light may still require a quantum treatment of the latter, as the measured signals can contain non-negligible contributions from the quantum vacuum of the fields that interact with the medium under study. We demonstrate that vacuum contributions can be expressed as a correction term to the \emph{classical} pump-probe signals, and it scales linearly with the intensity of the pump field. Contrary to popular belief, we show that these vacuum contributions may not be negligible, scaling \emph{quadratically} with the number of molecules $N_{mol}$ in the medium, and leading to the observation of superradiance in pump-probe experiments, provided that $N_{mol}$ is comparable to the number of photons in each pulse, and that a collinear configuration of the experiment is adopted.

The importance of our results is threefold. First, they explicitly show the quantum contribution contained in typical laser-spectroscopy signals, whose origin lies in the non-commutativity of the creation and annihilation operators that describe the optical fields; second, they provide a simple scheme for the experimental verification of the quantum nature of laser spectroscopy, which may be tested in current pump-probe experimental setups; third, they constitute a realistic example where the standard semiclassical treatment of light-matter interaction fails to describe it. Even though with different results and aims, previous related studies on the role of vacuum fluctuations in nonlinear spectroscopy have also been reported in Refs. \cite{martinez2005,bennett2014,glenn2015}.



\emph{Model.---}We start by considering the absorbing medium as a coupled dimer molecule, which is constructed by coupling two two-level chromophores (sites) $A$ and $B$ [Fig. 1 (a)] \cite{cho2009,joel_book,joel2011,joel2011-2}. The Hamiltonian of this molecular system is given by
\begin{equation}\label{Hamiltonian1}
\hat{H}_{0}/\hbar = \sum_{i=g,a,b,f} \omega_{i}\ket{i}\bra{i} + J\pare{\ket{a}\bra{b} + \ket{b}\bra{a}},
\end{equation}
where $\ket{g}$ is the electronic state where both chromophores are in their ground states, $\ket{a}$ and $\ket{b}$ are the electronic states where only one chromophore ($A$ or $B$, respectively) is excited, and $\ket{f}$ is the electronic state where both $A$ and $B$ are excited. $\omega_{i}$ is the energy of the $i$th electronic state, and $J$ is the coupling strength between the two chromophores. Notice that $\omega_{f} = \omega_{a}+\omega_{b}$, which is a statement of the lack of interaction between the two excitations in the doubly excited state.

\begin{figure}[t!]
\begin{center}
       \includegraphics[width=8.5cm]{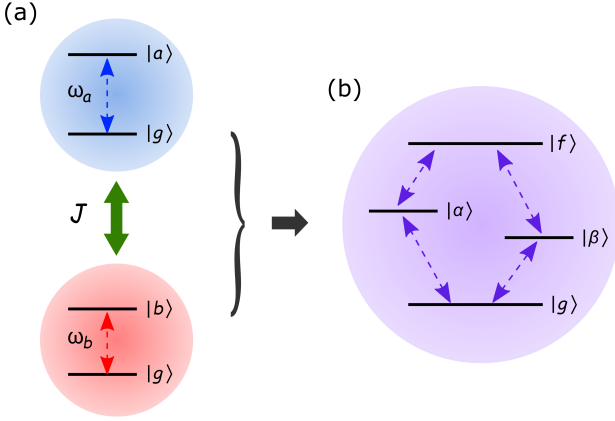}
\end{center}
\label{fig:levels} \caption{Energy level diagram of (a) two coupled chromophores, whose coupling strength is determined by the coefficient $J$, and (b) effective four-level system obtained upon diagonalization of the molecular Hamiltonian. Dashed arrows indicate allowed transitions.}
\end{figure}

Upon diagonalization of Eq. (\ref{Hamiltonian1}), the two-site system becomes an effective four-level system [shown in Fig. 1(b)] defined by the eigenvectors $\lbrace \ket{g},\ket{\alpha},\ket{\beta},\ket{f} \rbrace$ and eigenvalues $\lbrace \omega_{g},\omega_{\alpha},\omega_{\beta},\omega_{f} \rbrace$. The eigenvectors are given by $\ket{\alpha} = \cos\theta\ket{a} + \sin\theta\ket{b}$ and $\ket{\beta} = -\sin\theta\ket{a} + \cos\theta\ket{b}$; whereas the eigenvalues are $\omega_{\alpha} = \overline{\omega} + \delta\sec 2\theta$ and $\omega_{\beta} = \overline{\omega} - \delta\sec 2\theta$, where $\theta=\frac{1}{2}\arctan(J/\delta)$, $\delta = \pare{\omega_{a}-\omega_{b}}/2$, and $\overline{\omega} = \pare{\omega_{a} + \omega_{b}}/2$.

The interaction of the effective four-level system with an optical field can be described by the Hamiltonian in the interaction representation \cite{mollow1968,mukamel_book}

\begin{equation}
\hat{H}_{I}\pare{t} = - \bm{\mu}\pare{t}\cdot\cor{\hat{\bm{E}}^{\pare{+}}\pare{t} + \hat{\bm{E}}^{\pare{-}}\pare{t}},
\end{equation}
where $\bm{\mu}\pare{t}$ stands for the dipole-moment operator of the molecules and $\hat{\bm{E}}^{\pare{+}}\pare{t} = \cor{\hat{\bm{E}}^{\pare{-}}\pare{t}}^{*}$ for the positive-frequency part of the electric-field operator, which is given by $\hat{\bm{E}}^{\pare{+}}\pare{t} = \hat{\bm{E}}_{P}^{\pare{+}}\pare{t} + \hat{\bm{E}}_{P'}^{\pare{+}}\pare{t}$, with $\hat{\bm{E}}^{\pare{+}}_{P,P'}\pare{t}$ denoting the operators for the pump ($P$) and probe ($P'$) fields, defined by \cite{garrison_book}
\begin{equation}\label{quantum_field}
\hat{\bm{E}}^{\pare{+}}_{j}\pare{t} = \sum_{k_{j}}i\sqrt{\frac{\hbar\omega_{k_{j}}}{2\epsilon_{0}V}}\hat{a}_{k_{j}}\bm{e}_{k_{j}}\exp\cor{-i\omega_{k_{j}}\pare{t-t_{j}}},
\end{equation}
with $j=P,P'$. Here, $\epsilon_{0}$ is the vacuum permittivity, $V$ is the quantization volume, $\hat{a}_{k_{j}}$ is the annihilation operator of a photonic mode bearing a frequency $\omega_{k_{j}}$ and polarization $\bm{e}_{k_{j}}$, and $t_{j}$ is a time delay due to the propagation of the field.

\begin{figure}[t!]
\begin{center}
       \includegraphics[width=8.5cm]{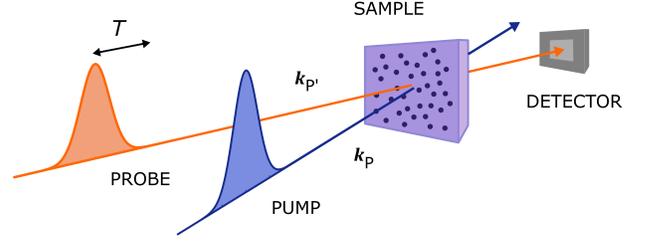}
\end{center}
\label{setup}
\caption{Pump-probe spectroscopy. A pump pulse $P$ excites the sample. After a waiting time $T$, the number of photons lost by a probe pulse $P'$ passing through that sample is measured.}
\end{figure}

In a typical pump-probe configuration [as the one depicted in Fig. (2)], a pump pulse $P$ traveling in the direction $\bm{k}_{P}$ excites the molecules of a sample from the ground state to their bright excited states via the dipole operator. After a waiting time $T$, a probe pulse $P'$ traveling in the $\bm{k}_{P'}$ direction passes through the sample, and the number of photons lost by this pulse is recorded. This measured pump-probe signal ($S_{PP'}$) is what allows us to characterize the excited and ground state non-equilibrium dynamics of a molecular system.


We can make use of second-order time-dependent perturbation theory to derive the expressions for the two-photon processes that contribute to a typical pump-probe signal, considering an arbitrary initial state of the pump and probe pulses (see section I of supplementary material for details). In what follows, we will present the solution to Eqs. (S1)-(S3) using multimode coherent states, which are the states that best describe the light-source used in typical experimental pump-probe spectroscopy.

\emph{Results.---}We consider the initial state of the optical field as composed by two laser pulses. This type of field can be well represented as \cite{brumer2012}
\begin{equation}
\ket{\varphi_{i}} = \ket{\llav{\alpha_{k}}}_{P}\ket{\llav{\alpha_{k}}}_{P'},
\label{initial-state}
\end{equation}
where $\ket{\llav{\alpha_{k}}}_{j}$ stands for a multimode coherent state defined by \cite{garrison_book,brumer2012,loudon_book}
\begin{equation}\label{multimode-coherent}
\ket{\llav{\alpha_{k}}}_{j} = \prod_{k} \ket{\alpha_{k}},
\end{equation}
with
\begin{equation}\label{coherent}
\ket{\alpha_{k}} = \exp\pare{-\abs{\alpha_{k}}^{2}/2}\sum_{N=0}^{\infty}\frac{\alpha_{k}^{N}}{N!}\pare{\hat{a}_{k}^{\dagger}}^{N}\ket{\text{vac}}.
\end{equation}
The parameter $\alpha_{k}$ defines the complex amplitude of the $k$th coherent state, whose squared norm represents the most likely number of photons carried by the corresponding mode, \emph{i.e.}, its intensity.

Now, by substituting Eq. (\ref{initial-state}) into Eqs. (S1)-(S3) it is easy to show that the typical pump-probe spectroscopy signal of the described effective four-level molecular system [as illustrated in Fig. 1(b)] is $S_{PP'}=S_{ESA}+S_{SE}+S_{GSB}$, where each of the terms is given by \cite{joel_book}

\begin{figure}[t!]
\begin{center}
       \includegraphics[width=7.5cm]{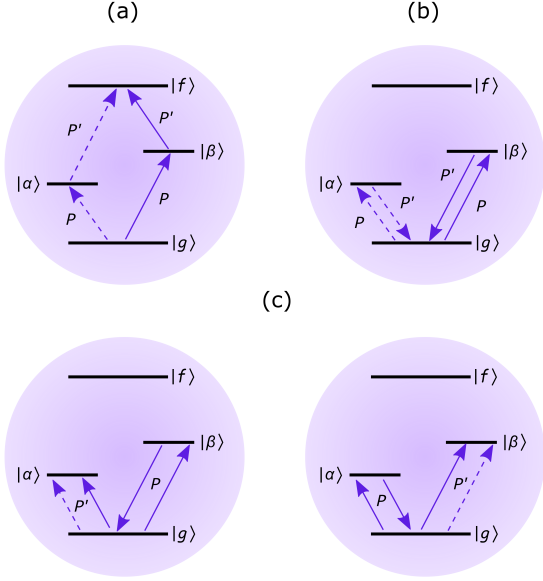}
\end{center}
\label{pathways}
\caption{Two-photon absorption pathways involved in the process of (a) excited-state absorption ($S_{\text{ESA}}$), (b) stimulated emission ($S_{\text{SE}}$) and (c) ground-state bleach ($S_{\text{GSB}}$). Bra and ket pathways are indicated by dashed and solid arrows, respectively.}
\end{figure}

\begin{eqnarray}
&S_{ESA}& = \abs{\Omega_{f\alpha}^{P'}\Omega_{\alpha g}^{P}e^{-i\omega_{\alpha}T} + \Omega_{f\beta}^{P'}\Omega_{\beta g}^{P}e^{-i\omega_{\beta}T} }^{2}, \label{Sesa}\\
&S_{SE}& = -\abs{\Omega_{g\alpha}^{\overline{P'}}\Omega_{\alpha g}^{P}e^{-i\omega_{\alpha}T} + \Omega_{g\beta}^{\overline{P'}}\Omega_{\beta g}^{P}e^{-i\omega_{\beta}T} }^{2} \nonumber \\
& & \hspace{4mm} -\left[ \abs{\Omega_{g\alpha}^{\text{vac},\overline{P'}}\Omega_{\alpha g}^{P} }^{2} + \abs{\Omega_{g\beta}^{\text{vac},\overline{P'}}\Omega_{\beta g}^{P} }^{2}\right. \nonumber \\
& & \hspace{4mm} +\left. 2\Delta_{\alpha\beta}\mathcal{R}\pare{\Omega_{g\alpha}^{\text{vac},\overline{P'}}\Omega_{\alpha g}^{P}\Omega_{g\beta}^{\overline{P}}\Omega_{\beta g}^{\text{vac},P'} }\right], \label{Sse}\\
&S_{GSB}& = -\pare{\abs{\Omega_{\alpha g}^{P}}^{2} + \abs{\Omega_{\beta g}^{P}}^{2}}\pare{\abs{\Omega_{\alpha g}^{P'}}^{2} + \abs{\Omega_{\beta g}^{P'}}^{2}},\label{Sgsb}
\end{eqnarray}
where $\Omega_{qn}^{j} = \frac{\mu_{qn}}{\hbar}\eta_{j}e^{-\frac{\sigma_{j}^{2}}{2}\pare{\omega_{q}-\omega_{n}-\omega_{j}^{0}}^{2}} = (\Omega_{nq}^{\overline{j}})^{*}$, with $n=\alpha,\beta$, and $q=g,f$. $\sigma_{j}$ is the time duration of the $j$th Gaussian pulse, characterized by an amplitude $\eta_{j} = \cor{\hbar\omega_{j}^{0}N_{j}\sqrt{\pi}\sigma_{j}/\pare{\epsilon_{0}cA}}^{1/2}$, which depends on the number of photons $N_{j}$ centered about the average frequency $\omega_{j}^{0}$. $\mu_{qn} = \bra{q}\bm{\mu}\cdot\bm{e}\ket{n}$ is a transition matrix element of the dipole-moment operator (see section II of the supplementary materials for details). In Eq. (\ref{Sse}), $\mathcal{R}\pare{\cdots}$ stands for the real part, and a window function such that $\Delta_{\alpha\beta} = 1$ if $\omega_{\beta} \in \cor{\omega_{\alpha}-\pi/(2\Gamma),\omega_{\beta}+\pi/(2\Gamma)}$, and $\Delta_{\alpha\beta} = 0$ otherwise (see section IB of supplementary materials). Notice that Eqs. (\ref{Sesa})-(\ref{Sgsb}) have been written in a suggestive manner in order to make a direct connection with Fig. (3). It coincides with the calculation where the pulses are considered classical \cite{joel_book}, with the curious exception of the additional contribution in the second and third lines of Eq. (\ref{Sse}). Ignoring the latter for the time being, excited-state absorption ($S_{ESA}$) and stimulated emission ($S_{SE}$) may be identified as double-slit-like processes, where the two-photon pathways interfere either in the doubly-excited [Fig. 3(a), Eq. (\ref{Sesa})] or the ground [Fig. 3(b), Eq. (\ref{Sse})] state, by consuming or yielding photons to the probe field, respectively. These interfering pathways give rise to population (which depend on transitions with $\ket{\alpha}$ or $\ket{\beta}$) and coherence terms (which depend on both types of transitions simultaneously, and modulate the signal with \emph{T}-dependent oscillatory quantum beats, where $T=t_{P'}-t_{P}$ is the waiting time). On the other hand, for the $S_{GSB}$ process [Fig. 3(c)], coherences between pathways involving $\ket{\alpha}$ and $\ket{\beta}$ do not occur because our simplified molecular model does not contain more than one vibronic level in the ground state, which are needed to produce interference effects; hence, only populations occur \cite{joel_book}. Therefore, in the model considered here, ground-state bleach can be understood as the decreased $P'$ absorption of the molecular system once $P$ has depleted some population from its ground state, and hence, can be computed as the (incoherent) composition of two absorption probabilities [see Eq. (\ref{Sgsb})]. In passing, an interesting limit is worth highlighting: when $J=0$, $\ket{\alpha}=\ket{a}$, $\ket{\beta}=\ket{b}$, $\mu_{fa}=\mu_{bg}$, and $\mu_{fb}=\mu_{ag}$. Under this circumstance, the molecular transitions become independent and quantum beats, characterized by the $T$-oscillatory terms in Eqs. (\ref{Sesa}) and (\ref{Sse}), vanish for $S_{PP'}$. Hence, the presence of quantum beats signals the availability of an energy transfer mechanism, represented by $J\neq 0$ in our case (see also Refs. \cite{joel2012,johson2014} for other scenarios with quantum beats based on vibronic coupling).

We are now ready to discuss the second and third lines of Eq. (\ref{Sse}). By analogy to $\Omega_{gq}^{\overline{P'}}$, we have defined $\Omega_{gq}^{\text{vac},\overline{P'}}=\frac{\mu_{gq}}{\hbar}\eta_{P'}^{\text{vac}}$, where $\eta_{P'}^{\text{vac}}=\cor{\hbar\omega_{P'}^{0}\Gamma/\pare{\epsilon_{0}cA}}^{1/2}$ is the single-photon electric field amplitude corresponding to vacuum fluctuations of the probe ($\Gamma$ is an average lifetime of the states $\ket{\alpha}$ and $\ket{\beta}$; a similar strategy invoking a finite linewidth for optical transitions was put forward in \cite{martinez2005}), and hence, should be regarded as a quantum correction to $S_{PP'}$. Notice that this contribution does not involve quantum beats because the creation and annihilation operators of photons commute unless they belong to the same mode ($[\hat{a}_{k},\hat{a}_{k'}^{\dagger}]=\delta_{kk'}$); quantum beats depend on the creation and annihilation of $P'$ photons of different color, and therefore, do not experience the mode vacua. However, double-slit static interferences survive between emission pathways involving $\ket{\alpha}$ and $\ket{\beta}$ as long as $\omega_{\alpha} \approx \omega_{\beta}$, and in particular, in the case where $J=0$ and $\omega_{\alpha} \approx \omega_{\beta}$, that is, even when the two chromophores are not interacting.


The origin of this quantum term is strictly analogous to the well-known quantum correction to spontaneous emission, where the rate does not scale classically as the number of photons in free-space $N$, but rather as $N+1$ \cite{schatz_book}. Similarly, our classical term in Eq. (\ref{Sse}) scales as the product of the number of photons in each pulse ($N_{P}N_{P'}$), while the quantum term scales linearly with the number of photons of the pump ($N_{P}$). This is the reason why in typical experiments, where light pulses contain a large number of photons, one may feel tempted to neglect the quantum term in Eq. (\ref{Sse}), thus regarding laser spectroscopy as a classical enterprise.  Notwithstanding, as we will show next, this assumption might not always be correct, particularly in the case where a collinear configuration of the pulses is adopted and, more importantly, where the absorbing medium contains a number of molecules similar to the number of photons in the probe field.

In typical pump-probe experiments, one deals with samples containing more than one molecule (single-molecule setups can in principle be carried out in fluorescence setups, see Refs. \cite{cina2008,lott2011,pachon2015}). This implies that in order to describe a realistic situation, we need to generalize our quantum calculation to a many-molecule case. To that end, we consider the simplest case in which the absorbing medium consists of a mixture of $N_{mol}$ dimers, where the singly and doubly excited states of the $i$th dimer are represented by $\ket{\alpha_{i}}$, $\ket{\beta_{j}}$, and $\ket{f_{j}}$, respectively. To simplify notation, we consider the case where $J\neq 0$ and $\Delta_{\alpha\beta}=0$. In this situation, by taking into account the spatially dependent phase of the fields interacting with the $i$th dimer at position $\bm{r}_i$, one can find that the pump-probe signal using multimode coherent states is given by (see section III of supplementary materials for details)

\begin{eqnarray}
&S_{PP'}& = N_{mol}\Bigg[ \sum_{n=\alpha,\beta}\abs{\Omega_{ng}^{P}}^{2}\pare{\abs{\Omega_{fn}^{P'}}^{2} - 2\abs{\Omega_{ng}^{P'}}^{2}}  \nonumber \\
&& \hspace{15mm} - \abs{\Omega_{\alpha g}^{P}}^{2}\abs{\Omega_{\beta g}^{P'}}^{2} - \abs{\Omega_{\beta g}^{P}}^{2}\abs{\Omega_{\alpha g}^{P'}}^{2} \Bigg] \nonumber \\
&& \hspace{2mm} + N_{mol}\cor{\Omega_{\alpha g}^{P}\Omega_{g\beta}^{\overline{P}}\Omega_{f\alpha}^{P'}\Omega_{\beta f}^{\overline{P'}}e^{-i\omega_{\alpha\beta}T} + \text{c.c.} } \nonumber \\
&& \hspace{2mm} - N_{mol}\cor{\Omega_{\alpha g}^{P}\Omega_{g\beta}^{\overline{P}}\Omega_{g\alpha}^{\overline{P'}}\Omega_{\beta g}^{P'}e^{-i\omega_{\alpha\beta}T} + \text{c.c.} } \nonumber \\
&& \hspace{2mm} - \abs{\sum_{j}e^{i\pare{\bm{k}_{P} - \bm{k}_{P'} }\cdot\bm{r}_{j}}}^{2}\sum_{n=\alpha,\beta}\abs{\Omega_{gn}^{\text{vac},\overline{P'}}\Omega_{ng}^{P}}^{2}.
\label{Spp-many}
\end{eqnarray}
In this expression, the classical contributions can be found in the first four rows, , and they simply amount to $N_{mol}$ times the classical contributions in Eqs. (\ref{Sesa})-(\ref{Sgsb}). The quantum contribution corresponds to the last term, which vanishes for $\bm{k}_{P} \neq \bm{k}_{P'}$, but reduces to $-N_{mol}^{2}\sum_{i=\alpha,\beta}\abs{\Omega_{gi}^{\text{vac},\overline{P'}}\Omega_{ig}^{P}}^{2}$ for the collinear configuration, where $\bm{k}_{P} = \bm{k}_{P'}$. This quadratic dependence on $N_{mol}$ is the footprint of superradiance \cite{dicke1954}, which emerges from the $N_{mol}^{2}$ possibilities for choosing a pair of chromophores $i$, $j$ which can emit a photon of frequency $\omega_{\alpha}$ from their respective $\ket{\alpha_{i}}$ and $\ket{\alpha_{j}}$ (or of frequency $\omega_{\beta}$ from $\ket{\beta_{i}}$ and $\ket{\beta_{j}}$) states. This fascinating result, whose physical origin is the same as that of the $\Delta_{\alpha\beta}$ term in Eq. (\ref{Sse}), suggests that vacuum contributions from the optical fields produce superradiant signals that could be detected in current pump-probe experimental setups, provided that $N_{mol}\Gamma \simeq N_{P'}\sigma_{P'}$, which makes the quantum and classical terms in Eq. (\ref{Spp-many}) comparable in magnitude. For large $N_{mol}$, the superradiant term dominates the signal and the intensity of the quantum beats (which belong to the classical term) becomes negligible compared to the static part of the signal. By taking ratios of these intensities as well as monitoring the signal dependence with respect to $N_{mol}$, one could conceivably detect this predicted classical to quantum transition.


Experimentally, one might proceed with an ensemble of bright organic dimer chromophores. In particular, one could use a solution of special-pair dimers extracted from the reaction of center of the photosynthetic purple bacterium \emph{Rhodobacter sphaeroides}. We have chosen this dimer because of its well-defined absorption spectrum and, more importantly, because it has been previously used in experimental nonlinear spectroscopy \cite{lee2007}. However, it is important to remark that the onset of the described superradiant effect does not depend on the strength of the optical transitions as long as they are bright enough to be experimentally detected.

Optical excitation of the sample could be carried out by means of transform-limited optical pulses ($\sigma_{j} \sim 20$ fs), centered at a wavelength of $775$ nm, carrying an energy per pulse of $5$ nJ ($\sim 1.9 \times 10^{10}$ photons per pulse), which corresponds to energies typically used in pump-probe experiments \cite{engel2007}. By using these parameters, one can find that the spectral width of the pulses are $\sim 44$ nm, which perfectly overlaps with the absorption lines of the dimer: $750$ nm and $800$ nm \cite{lee2007}. Finally, the sample's concentration is given by $C = N_{mol}/\pare{\mathcal{N}_{A}V}$, where $\mathcal{N}_{A}$ is the Avogadro constant, and the coherent laser volume, $V=AL$, is defined by the transversal area of the pulse $A=\pi D^{2}/4$, and its coherence length $L=c\sigma_{j}=6 \mu$m. By considering a typical laser spot diameter of $D=70\;\mu$m \cite{engel2007}, we can choose a concentration of $C \sim 1.4\;$mM that yields a number of molecules within the coherent volume of the laser of $N_{mol} \sim 1.9 \times 10^{10}$, which coincides with the number of photons contained in each pulse. Assuming that $\Gamma \simeq 400 fs$, which is a characteristic dephasing time between excited states in these systems \cite{lee2007}, we see that the superradiant quantum contribution becomes $\sim 20$ times larger than the classical one, so one could even decrease the concentration significantly. Given that the parameters have been taken from experimental studies, we believe that the predicted effects are well within experimental reach.


\emph{Conclusions.---}In summary, we have shown that, even though the standard semiclassical treatment of laser spectroscopy works well in a large number of cases, their signals are inherently \textit{quantum}, as they contain information about the quantum vacuum of the optical fields that interact with the absorbing medium. Interestingly, we have shown that quantum contributions may lead to the observation of superradiant phenomena, provided that the number of molecules in the medium is comparable to the number of photons in each pulse, and that a collinear configuration of the experiment is adopted. Furthermore, we have provided a simple scheme for the experimental verification of the quantum nature of laser spectroscopy, which may be tested in current pump-probe experimental setups. Finally, this paper highlights an important situation where the semiclassical treatment of light-matter interaction fails, and where a proper quantum treatment of light and matter becomes fundamental. We anticipate scenarios where these features might yield novel capabilities to nonlinear spectroscopies.

JYZ acknowledges discussions on the topic with Semion K. Saikin, Leonardo Pach\'{o}n, and Kochise Bennett. RJLM acknowledges postdoctoral financial support from the University of California Institute for Mexico and the United States (UC MEXUS). RJLM, ZH, and JYZ acknowledge generous startup funds from UCSD.

\clearpage

\begin{widetext}

\begin{flushleft}
\Large{\textbf{Supplementary material:\vspace{2mm}}}\\
\Large{\textbf{A genuinely quantum effect in laser spectroscopy: vacuum and its induced superradiance}}
\end{flushleft}
\begin{flushleft}
\normalsize{Roberto de J. Le\'{o}n-Montiel, Zixuan Hu, and Joel Yuen-Zhou}
\end{flushleft}
\vspace{0.5cm}

In this supplementary material, we present (i) the general form of the two-photon processes that contribute to a typical pump-probe signal, (ii) the derivation of Gaussian optical pulses from the complex amplitudes of multimode coherent states, and (iii) the quantum-mechanical calculation of pump-probe signals for many-molecule samples.

\section*{\normalsize{I. Two-photon processes for an arbitrary initial matter and light state}}

As discussed in the main text, a typical pump-probe signal can be written as a sum of three two-photon processes: excited-state absorption ($S_{\text{ESA}}$), stimulated emission ($S_{\text{SE}}$), and ground-state bleach ($S_{\text{GSB}}$). Assuming that the molecular system is in its ground state $\ket{g}$ and that the sample is illuminated by an arbitrary initial optical field $\ket{\varphi_{i}}$ (which includes both the pump $P$ and probe $P'$ fields), one can make use of second-order time-dependent perturbation theory to show that $S_{PP'}=S_{ESA}+S_{SE}+S_{GSB}$, where each of the terms is given by wavepacket overlaps \cite{joel_book}
\begin{align}
S_{\text{ESA}} &= \langle\Psi_{+P+P'}\ket{\Psi_{+P+P'}} , \label{esa}\tag{S1} \\
S_{\text{SE}}  &= -\langle\Psi_{+P-P'}\ket{\Psi_{+P-P'}}, \label{se}\tag{S2}\\
S_{\text{GSB}} &= 2\mathcal{R}\llav{\langle \Psi_{+P-P+P'}\vert\Psi_{+P'} \rangle}. \label{gsb} \tag{S3}
\end{align}
$\mathcal{R}\llav{\cdots}$ stands for the real part, and

\begin{equation}
\ket{\Psi_{+P'}} = \frac{i}{\hbar}\sum_{m=\alpha,\beta} \mu_{mg}\int_{-\infty}^{\infty} dt_{1}e^{-i\pare{\omega_{g}-\omega_{m}}t_{1}}\hat{E}_{P'}^{\pare{+}}\pare{t_{1}}\ket{m}\ket{\varphi_{i}},
\tag{S4}
\end{equation}

\begin{equation}\label{esa-ket}
\begin{split}
\ket{\Psi_{+P+P'}} = -\frac{1}{\hbar^{2}}\sum_{m=\alpha,\beta}\mu_{fm}\mu_{mg}&\int_{-\infty}^{\infty}dt_{2}\int_{-\infty}^{\infty}dt_{1}e^{-i\pare{\omega_{m}-\omega_{f}}t_{2}}e^{-i\pare{\omega_{g}-\omega_{m}}t_{1}}\\
& \times \hat{E}_{P'}^{\pare{+}}\pare{t_{2}}\hat{E}_{P}^{\pare{+}}\pare{t_{1}}\ket{f}\ket{\varphi_{i}},
\end{split}
\tag{S5}
\end{equation}

\begin{equation}\label{se-ket}
\begin{split}
\ket{\Psi_{+P-P'}} = -\frac{1}{\hbar^{2}}\sum_{m=\alpha,\beta}\mu_{gm}\mu_{mg}&\int_{-\infty}^{\infty}dt_{2}\int_{-\infty}^{\infty}dt_{1}e^{-i\pare{\omega_{m}-\omega_{g}}t_{2}}e^{-i\pare{\omega_{g}-\omega_{m}}t_{1}}\\
& \times \hat{E}_{P'}^{\pare{-}}\pare{t_{2}}\hat{E}_{P}^{\pare{+}}\pare{t_{1}}\ket{g}\ket{\varphi_{i}},
\end{split}
\tag{S6}
\end{equation}

\begin{equation}\label{gsb-ket}
\begin{split}
\ket{\Psi_{+P-P+P'}} = -\frac{i}{\hbar^{3}}\sum_{m,n=\alpha,\beta} \mu_{mg}\mu_{gn}\mu_{ng}&\int_{-\infty}^{\infty} dt_{3} \int_{-\infty}^{\infty}dt_{2}\int_{-\infty}^{t_{2}}dt_{1}e^{-i\pare{\omega_{g}-\omega_{m}}t_{3}}e^{-i\pare{\omega_{n}-\omega_{g}}t_{2}} \\
& \times e^{-i\pare{\omega_{g}-\omega_{n}}t_{1}}\hat{E}_{P'}^{\pare{+}}\pare{t_{3}} \hat{E}_{P}^{\pare{-}}\pare{t_{2}}\hat{E}_{P}^{\pare{+}}\pare{t_{1}} \ket{m}\ket{\varphi_{i}}.
\end{split}
\tag{S7}
\end{equation}
where $\mu_{qn} = \bra{q}\bm{\mu}\cdot\bm{e}\ket{n}$, with $q=g,f$ are the transition dipole moments and $\hat{E}_{P,P'}^{\pare{\pm}}\pare{t}$ are defined by Eq. (3) in the main text. It is important to highlight that, in writing Eqs. (\ref{esa-ket}) and (\ref{se-ket}), we have assumed that the pump and probe pulses do not overlap in time, which allows us to extend most integrals to be from $-\infty$ to $\infty$. This is not the case for Eq. (\ref{gsb-ket}), where nested integrals cannot be decoupled since they involve a second order perturbation with the same pulse \cite{joel_book}. In what follows, we provide a detailed derivation for each of the process described by Eqs. (S1)-(S3). Special attention will be paid to stimulated emission [Eq. (S2)], as it is the process from which quantum vacuum contributions emerge.

\subsection*{\normalsize{A. Excited-state absorption}}

To compute the $S_{ESA}$ signal, we start by substituting Eq. (S5) into Eq. (S1) to obtain
\begin{equation}
\begin{split}
S_{ESA}=&\frac{1}{\hbar^{4}}\int_{-\infty}^{\infty}dt_{2}\int_{-\infty}^{\infty}dt_{1}\int_{-\infty}^{\infty}dt'_{2}\int_{-\infty}^{\infty}dt'_{1}
\sum_{m=\alpha,\beta}\mu_{fm}\mu_{mg}e^{-i\pare{\omega_{m}-\omega_{f}}t_{2}}e^{-i\pare{\omega_{g}-\omega_{m}}t_{1}} \\
&\times \sum_{n=\alpha,\beta}\mu_{fn}^{*}\mu_{ng}^{*}e^{i\pare{\omega_{n}-\omega_{f}}t'_{2}}e^{i\pare{\omega_{g}-\omega_{n}}t'_{1}} \bra{\varphi_{i}}\hat{E}_{P}^{\pare{-}}\pare{t'_{1}}\hat{E}_{P'}^{\pare{-}}\pare{t'_{2}}\hat{E}_{P'}^{\pare{+}}\pare{t_{2}}\hat{E}_{P}^{\pare{+}}\pare{t_{1}}\ket{\varphi_{i}}.
\end{split}
\tag{S8}
\end{equation}
Now, by considering the initial state of the field as a multimode coherent state, defined by Eq. (4) in the main text, we can write the term involving the field operators as
\begin{equation}
\begin{split}
\bra{\varphi_{i}}\hat{E}_{P}^{\pare{-}}\pare{t'_{1}}\hat{E}_{P'}^{\pare{-}}\pare{t'_{2}}\hat{E}_{P'}^{\pare{+}}\pare{t_{2}}\hat{E}_{P}^{\pare{+}}\pare{t_{1}}\ket{\varphi_{i}}=&\bra{\llav{\alpha}}_{P}\hat{E}_{P}^{\pare{-}}\pare{t'_{1}}\hat{E}_{P}^{\pare{+}}\pare{t_{1}}\ket{\llav{\alpha}}_{P} \\
& \times \bra{\llav{\alpha}}_{P'}\hat{E}_{P'}^{\pare{-}}\pare{t'_{2}}\hat{E}_{P'}^{\pare{+}}\pare{t_{2}}\ket{\llav{\alpha}}_{P'},
\end{split}
\tag{S9}
\end{equation}
where we have factorized the terms corresponding to the operators for the pump and probe fields, respectively (this cannot be done in the case of entangled photons \cite{roslyak2009,roslyak2009-2,roberto2013}). By making use of Eq. (3) of the main manuscript, we can show that
\begin{align}
\bra{\llav{\alpha}}_{P}\hat{E}_{P}^{\pare{-}}\pare{t'_{1}}\hat{E}_{P}^{\pare{+}}\pare{t_{1}}\ket{\llav{\alpha}}_{P} &=
\bra{\llav{\alpha}}_{P} \sum_{l}\sqrt{\frac{\hbar\omega_{l}}{2\epsilon_{0}V}}\hat{a}_{l}^{\dagger}e^{i\omega_{l}\pare{t'_{1}-t_{P}}} \sum_{k}\sqrt{\frac{\hbar\omega_{k}}{2\epsilon_{0}V}}\hat{a}_{k}e^{-i\omega_{k}\pare{t_{1}-t_{P}}} \ket{\llav{\alpha}}_{P}, \nonumber \\
&= \sum_{l}\sum_{k}\sqrt{\frac{\hbar\omega_{l}}{2\epsilon_{0}V}}e^{i\omega_{l}\pare{t'_{1}-t_{P}}} \sqrt{\frac{\hbar\omega_{k}}{2\epsilon_{0}V}}e^{-i\omega_{k}\pare{t_{1}-t_{P}}}\bra{\llav{\alpha}}_{P} \hat{a}_{l}^{\dagger} \hat{a}_{k}\ket{\llav{\alpha}}_{P}, \nonumber \\
&=\sum_{l}\sum_{k}\sqrt{\frac{\hbar\omega_{l}}{2\epsilon_{0}V}}e^{i\omega_{l}\pare{t'_{1}-t_{P}}} \sqrt{\frac{\hbar\omega_{k}}{2\epsilon_{0}V}}e^{-i\omega_{k}\pare{t_{1}-t_{P}}}\bra{\llav{\alpha}}_{P} \alpha_{l}^{*} \alpha_{k}\ket{\llav{\alpha}}_{P}, \nonumber \\
&=\sum_{l}\sqrt{\frac{\hbar\omega_{l}}{2\epsilon_{0}V}}\alpha_{l}^{*}e^{i\omega_{l}\pare{t'_{1}-t_{P}}} \sum_{k}\sqrt{\frac{\hbar\omega_{k}}{2\epsilon_{0}V}}\alpha_{k}e^{-i\omega_{k}\pare{t_{1}-t_{P}}}, \nonumber \\
&=\varepsilon_{P}^{*}\pare{t'_{1}-t_{P}}\varepsilon_{P}\pare{t_{1}-t_{P}},\tag{S10}
\end{align}
where we identify $\varepsilon_{P}\pare{t} = \sum_{k}\sqrt{\frac{\hbar\omega_{k}}{2\epsilon_{0}V}}\alpha_{k}e^{-i\omega_{k}t}$ as the temporal shape of the pump pulses \cite{joel_book}. Similarly, for the probe field, we obtain
\begin{equation}
\bra{\llav{\alpha}}_{P'}\hat{E}_{P'}^{\pare{-}}\pare{t'_{2}}\hat{E}_{P'}^{\pare{+}}\pare{t_{2}}\ket{\llav{\alpha}}_{P'} = \varepsilon_{P'}^{*}\pare{t'_{2}-t_{P'}}\varepsilon_{P'}\pare{t_{2}-t_{P'}}.
\tag{S11}
\end{equation}
Finally, by assuming that the temporal shapes of the optical pulses are described by Gaussian functions (see section II for details on the Gaussian pulse approximation), we can substitute Eqs. (S10)-S(11) into Eq. (S8) to obtain, via evaluation of Gaussian Fourier integrals, the results presented in Eq. (7) of the main text (see also \cite{joel_book} for further details).

\subsection*{\normalsize{B. Stimulated emission}}

From all the processes described in this work, $S_{SE}$ is the most important one, as it is the term that contains information about the vacuum fluctuations of the probe field, which ultimately leads to the quantum correction of $S_{PP'}$. To compute this term, we substitute Eq. (S6) into (S2) to obtain
\begin{equation}
\begin{split}
S_{SE} = &-\frac{1}{\hbar^{4}}\int_{-\infty}^{\infty}dt_{2}\int_{-\infty}^{\infty}dt_{1}\int_{-\infty}^{\infty}dt'_{2}\int_{-\infty}^{\infty}dt'_{1}
\sum_{m=\alpha,\beta}\mu_{gm}\mu_{mg}e^{-i\pare{\omega_{m}-\omega_{g}}t_{2}}e^{-i\pare{\omega_{g}-\omega_{m}}t_{1}} \\
&\times \sum_{n=\alpha,\beta}\mu_{gn}^{*}\mu_{ng}^{*}e^{i\pare{\omega_{n}-\omega_{g}}t'_{2}}e^{i\pare{\omega_{g}-\omega_{n}}t'_{1}} \bra{\varphi_{i}}\hat{E}_{P}^{\pare{-}}\pare{t'_{1}}\hat{E}_{P'}^{\pare{+}}\pare{t'_{2}}\hat{E}_{P'}^{\pare{-}}\pare{t_{2}}\hat{E}_{P}^{\pare{+}}\pare{t_{1}}\ket{\varphi_{i}}.
\end{split}
\tag{S12}
\end{equation}
Following the same procedure as in the previous subsection, we substitute Eq. (4) of the main text into Eq. (S12), so the term involving field operators can be written as
\begin{equation}
\begin{split}
\bra{\varphi_{i}}\hat{E}_{P}^{\pare{-}}\pare{t'_{1}}\hat{E}_{P'}^{\pare{-}}\pare{t'_{2}}\hat{E}_{P'}^{\pare{+}}\pare{t_{2}}\hat{E}_{P}^{\pare{+}}\pare{t_{1}}\ket{\varphi_{i}}=&\bra{\llav{\alpha}}_{P}\hat{E}_{P}^{\pare{-}}\pare{t'_{1}}\hat{E}_{P}^{\pare{+}}\pare{t_{1}}\ket{\llav{\alpha}}_{P} \\
& \times \bra{\llav{\alpha}}_{P'}\hat{E}_{P'}^{\pare{-}}\pare{t'_{2}}\hat{E}_{P'}^{\pare{+}}\pare{t_{2}}\ket{\llav{\alpha}}_{P'}.
\end{split}
\tag{S13}
\end{equation}
Notice that the term for the pump field is the same as in Eq. (S10), so we now focus only on the term involving the probe field and write
\begin{align}
\bra{\llav{\alpha}}_{P'}\hat{E}_{P'}^{\pare{-}}\pare{t'_{2}}\hat{E}_{P'}^{\pare{+}}\pare{t_{2}}\ket{\llav{\alpha}}_{P'} &=
\bra{\llav{\alpha}}_{P'} \sum_{l}\sqrt{\frac{\hbar\omega_{l}}{2\epsilon_{0}V}}\hat{a}_{l}e^{-i\omega_{l}\pare{t'_{2}-t_{P'}}} \sum_{k}\sqrt{\frac{\hbar\omega_{k}}{2\epsilon_{0}V}}\hat{a}_{k}^{\dagger}e^{i\omega_{k}\pare{t_{2}-t_{P'}}} \ket{\llav{\alpha}}_{P'}, \nonumber \\
&= \sum_{l}\sum_{k}\sqrt{\frac{\hbar\omega_{l}}{2\epsilon_{0}V}}e^{-i\omega_{l}\pare{t'_{2}-t_{P'}}} \sqrt{\frac{\hbar\omega_{k}}{2\epsilon_{0}V}}e^{i\omega_{k}\pare{t_{2}-t_{P'}}}\bra{\llav{\alpha}}_{P'} \hat{a}_{l}\hat{a}_{k}^{\dagger}\ket{\llav{\alpha}}_{P'}. \tag{S14}
\end{align}
Note that evaluation of Eq. (S14) is not straightforward unless we write the product of the annihilation and creation operators in normal ordering. To this end, we make use of the commutator
\begin{equation}
\cor{\hat{a}_{l},\hat{a}_{k}^{\dagger}} = \hat{a}_{l}\hat{a}_{k}^{\dagger} - \hat{a}_{k}^{\dagger}\hat{a}_{l} = \delta_{l,k},
\tag{S15}
\end{equation}
so Eq. (S14) takes the form
\begin{equation}
\bra{\llav{\alpha}}_{P'}\hat{E}_{P'}^{\pare{-}}\pare{t'_{2}}\hat{E}_{P'}^{\pare{+}}\pare{t_{2}}\ket{\llav{\alpha}}_{P'} = \varepsilon_{P'}\pare{t'_{2}-t_{P'}}\varepsilon_{P'}^{*}\pare{t_{2}-t_{P'}} + \sum_{k}\frac{\hbar\omega_{k}}{2\epsilon_{0}V}e^{-i\omega_{k}\pare{t'_{2}-t_{2}}}.
\tag{S16}
\end{equation}
Let us look at the two terms in Eq. (S16): the first corresponds to the $S_{SE}$ signal obtained when considering the pump and probe pulses as classical fields; whereas the second one is the term that accounts for the inherent quantum properties of the coherent-state probe field. This second term ultimately leads to the signal that we referred to as the quantum correction to $S_{PP'}$. For the sake of clarity, in what follows, we present a detailed derivation of this correction term.

We start by substituting Eqs. (S13), together with Eqs. (S14) and (S15), into Eq. (S12) and focus on the term where the quantum contribution appears
\begin{equation}
\begin{split}
\text{Correction} = -&\frac{1}{\hbar^{4}}\int_{-\infty}^{\infty}dt_{2}\int_{-\infty}^{\infty}dt_{1}\int_{-\infty}^{\infty}dt'_{2}\int_{-\infty}^{\infty}dt'_{1}\sum_{m=\alpha,\beta}\mu_{gm}\mu_{mg}e^{-i\pare{\omega_{m}-\omega_{g}}t_{2}}e^{-i\pare{\omega_{g}-\omega_{m}}t_{1}} \\
&\times \sum_{n=\alpha,\beta}\mu_{gn}^{*}\mu_{ng}^{*}e^{i\pare{\omega_{n}-\omega_{g}}t'_{2}}e^{i\pare{\omega_{g}-\omega_{n}}t'_{1}}\varepsilon_{P}^{*}\pare{t'_{1}-t_{P}}\varepsilon_{P}\pare{t_{1}-t_{P}}\sum_{k}\frac{\hbar\omega_{k}}{2\epsilon_{0}V}e^{-i\omega_{k}\pare{t'_{2}-t_{2}}}. \end{split}
\tag{S17}
\end{equation}
Notice that the sum in Eq. (S17) can be simplified by assuming that the bandwidth of the pulses are much smaller than their central frequencies, so the frequency of all the $k$-modes are approximately the same as the central frequency of the pulse. In this situation, we can write the sum as
\begin{equation}
\sum_{k}\frac{\hbar\omega_{k}}{2\epsilon_{0}V}e^{-i\omega_{k}\pare{t'_{2}-t_{2}}} \simeq \frac{\hbar\omega_{P'}^{0}}{2\epsilon_{0}V}\sum_{k}e^{-i\omega_{k}\pare{t'_{2}-t_{2}}} = \frac{\hbar\omega_{P'}^{0}}{2\epsilon_{0}cA}\delta\pare{t'_{2}-t_{2}},
\tag{S18}
\end{equation}
where the last term is obtained by making use of the continuous representation of the sum over the $k$ modes \cite{blow1990}.

Now, by substituting Eq. (S18) into (S17), we can write
\begin{align}
\text{Correction} =-&\frac{1}{\hbar^{4}}\int_{-\infty}^{\infty}dt_{2}\int_{-\infty}^{\infty}dt_{1}\int_{-\infty}^{\infty}dt'_{2}\int_{-\infty}^{\infty}dt'_{1}\sum_{m,n=\alpha,\beta}\mu_{gm}\mu_{mg}\mu_{gn}^{*}\mu_{ng}^{*}e^{-i\pare{\omega_{m}-\omega_{g}}t_{2}}e^{-i\pare{\omega_{g}-\omega_{m}}t_{1}} \nonumber \\
& \times e^{i\pare{\omega_{n}-\omega_{g}}t'_{2}}e^{i\pare{\omega_{g}-\omega_{n}}t'_{1}}\varepsilon_{P}^{*}\pare{t'_{1}-t_{P}}\varepsilon_{P}\pare{t_{1}-t_{P}}\frac{\hbar\omega_{P'}^{0}}{2\epsilon_{0}cA}\delta\pare{t'_{2}-t_{2}}, \nonumber \\
=-&\frac{1}{\hbar^{4}}\int_{-\infty}^{\infty}dt_{2}\int_{-\infty}^{\infty}dt_{1}\int_{-\infty}^{\infty}dt'_{1}\sum_{m,n=\alpha,\beta}\mu_{gm}\mu_{mg}\mu_{gn}^{*}\mu_{ng}^{*}e^{-i\pare{\omega_{g}-\omega_{m}}t_{1}}e^{i\pare{\omega_{g}-\omega_{n}}t'_{1}}e^{-i\pare{\omega_{m}-\omega_{n}}t_{2}} \nonumber \\
& \times \varepsilon_{P}^{*}\pare{t'_{1}-t_{P}}\varepsilon_{P}\pare{t_{1}-t_{P}}\frac{\hbar\omega_{P'}^{0}}{2\epsilon_{0}cA}.
\tag{S19}
\end{align}
Notice that, in Eq. (S19), the integral $\int_{-\infty}^{\infty}dt_{2}\exp\cor{-i\pare{\omega_{m}-\omega_{n}}t_{2}} = 2\pi\delta\pare{\omega_{m}-\omega_{n}}$ causes the Correction to diverge for $\omega_{m}=\omega_{n}$. In reality, this integral is quenched by the finite lifetime of the excited states. Regularizing the integrand by multiplying it by $\exp\pare{-\abs{t_{2}}/\Gamma}$,
\begin{equation}
\begin{split}
\int_{\infty}^{\infty}dt_{2}e^{-i\pare{\omega_{m}-\omega_{n}}t_{2}} & \rightarrow  \int_{-\infty}^{\infty} dt_{2}e^{-i\pare{\omega_{m}-\omega_{n}}t_{2}}e^{-\abs{t_{2}}/\Gamma} \\
& = \frac{2/\Gamma}{\pare{1/\Gamma}^{2}+\pare{\omega_{m}-\omega_{n}}^{2}}\simeq 2\Gamma,
\end{split}
\tag{S20}
\end{equation}
where $\Gamma$ is the lifetime of the $\ket{m}\bra{n}$ coherence (or if $\ket{m}=\ket{n}$, the population lifetime). Here, we have assumed that the Lorentzian function can be approximated as a discrete window function where $\Delta_{mn} = 1$ if $\omega_{m}\in \cor{\omega_{n}-\pi/(2\Gamma),\omega_{n}+\pi/(2\Gamma)}$ and $\Delta_{mn}=0$ otherwise (this approximation guarantees that the integrals of both the Lorentzian and the discrete window function with respect to $\omega_{m}$ remain equal). Physically, this window function says that the quasidegenerate transitions can interfere via quantum vacuum fluctuations. Hence, not only are population terms ($\ket{m}=\ket{n}$) affected by vacuum, but also coherences ($\ket{m}\neq\ket{n}$) as long as $\omega_{m}\approx \omega_{n}$. Owing to the finite lifetime of the transitions, strict degeneracy is not required, but only a quasidegeneracy up to an energy window of order of $\Gamma^{-1}$. As explained in the main text, transitions which are not quasidegenerate in the aforementioned sense do not interfere via quantum vacuum fluctuations, as they depend on the creation and annihilation of $P'$ photons of different color, and therefore, do not experience the noncommutativity of the fields $\cor{\hat{a}_{k},\hat{a}_{k'}^{\dagger}}$. From here, it immediately follows that quantum beats are unaffected by these vacuum fluctuations.

Then, by substituting Eq. (S20) into Eq. (S19) we obtain
\begin{equation}
\begin{split}
\text{Correction} = -&\frac{1}{\hbar^{4}}\pare{\eta_{P'}^{\text{vac}}}^{2}\sum_{m,n=\alpha,\beta}\mu_{gm}\mu_{mg}\mu_{gn}^{*}\mu_{ng}^{*}\int_{-\infty}^{\infty}dt_{1}\varepsilon_{P}\pare{t_{1}-t_{P}}e^{-i\pare{\omega_{g}-\omega_{m}}t_{1}} \nonumber \\
& \hspace{35mm} \times \int_{-\infty}^{\infty}dt'_{1}\varepsilon_{P}^{*}\pare{t'_{1}-t_{P}}e^{i\pare{\omega_{g}-\omega_{n}}t'_{1}}.
\end{split}
\tag{S21}
\end{equation}
Using the approximation of the discrete window function in Eq. (S20), we have defined $\eta_{P'}^{\text{vac}}=\cor{\hbar\omega_{P'}^{0}\Gamma/(\epsilon_{0}cA)}^{1/2}$ as the vacuum single-photon amplitude, by analogy to the classical pulse amplitude $\eta_{j}$ (see main text as well as Eq. (S29)]. Here, the decoherence time $\Gamma$ plays the same role for the vacuum correction electric field amplitude as the pulse duration $\sqrt(pi)\sigma_{j}$ does for the classical part of the signal.

Finally, by assuming the temporal shapes of the optical pulses as Gaussian functions, we solve Eq. (S21) to obtain the results presented in Eq. (8) of the main text.

\subsection*{\normalsize{C. Ground-state bleach}}

Finally, we obtain the expression for the ground-state bleach signal. To that end, we make use of Eqs. (S4) and (S7) to write Eq. (S3) as
\begin{equation}
\begin{split}
S_{GSB} = & 2\mathcal{R}\Bigg\{ -\frac{1}{\hbar^{4}}\int_{-\infty}^{\infty}dt_{4}\int_{-\infty}^{\infty}dt_{3}\int_{-\infty}^{\infty}dt_{2}\int_{-\infty}^{t_{2}}dt_{1}
\sum_{m=\alpha,\beta}\mu_{gm}\mu_{mg}e^{-i\pare{\omega_{g}-\omega_{m}}t_{4}}e^{-i\pare{\omega_{m}-\omega_{g}}t_{3}} \\
&\hspace{7mm} \times \sum_{n=\alpha,\beta}\mu_{gn}^{*}\mu_{ng}^{*}e^{i\pare{\omega_{n}-\omega_{g}}t_{2}}e^{i\pare{\omega_{g}-\omega_{n}}t_{1}} \bra{\varphi_{i}}\hat{E}_{P}^{\pare{-}}\pare{t_{1}}\hat{E}_{P}^{\pare{+}}\pare{t_{2}}\hat{E}_{P'}^{\pare{-}}\pare{t_{3}}\hat{E}_{P'}^{\pare{+}}\pare{t_{4}}\ket{\varphi_{i}}\Bigg\}.
\end{split}
\tag{S22}
\end{equation}
Notice that the term involving the field interactions is the same as the one obtained for the $S_{ESA}$ signal [Eq. (S9)]. Therefore, we can use Eqs. (S10) and (S11) to solve Eq. (S22) and find the expression shown in Eq. (9) of the main text.

%
%

\section*{\normalsize{II. Gaussian pulse approximation}}

Throughout the manuscript, we have assumed that the optical pulses are described by Gaussian functions, however no explicit relationship between those pulses and their underlying quantum states was presented. In this section, we present the method for constructing Gaussian pulses from the complex amplitude of multimode coherent states, as well as the derivation of the field amplitudes, which is important in our work, as it explicitly contains information about the number of photons contained in each pulse.

For the sake of simplicity, let us consider the expression for the one-photon excited state [Eq. (S4)], and substitute Eq. (3) of the main text, to obtain
\begin{equation}
\ket{\Psi_{+P}} = \frac{i}{\hbar}\sum_{m=\alpha,\beta} \mu_{mg}\int_{-\infty}^{\infty} dt_{1}e^{-i\pare{\omega_{g}-\omega_{m}}t_{1}}\sum_{k}i\sqrt{\frac{\hbar\omega_{k}}{2\epsilon_{0}V}}\alpha_{k}e^{-i\omega_{k}\pare{t-t_{P}}}\ket{m}\ket{\llav{\alpha}}_{P}.
\tag{S23}
\label{one-photon}
\end{equation}
As discussed in the previous section, we identify the temporal shape of the pump field as
\begin{equation}
\varepsilon_{P}\pare{t-t_{P}} = \int d\omega_{k}\sqrt{\frac{\hbar\omega_{k}}{4\pi\epsilon_{0}cA}}\alpha_{k}\pare{\omega_{k}}e^{-i\omega_{k}\pare{t-t_{P}}},
\tag{S24}
\label{gaussian-pulses}
\end{equation}
where $c$ is the speed of light and $A$ is the effective area of the field. Notice that, in writing Eq. (\ref{gaussian-pulses}), we have used the continous-mode representation of Eq. (3), as described in Ref. \cite{blow1990}.

We now proceed by taking the complex amplitude of the $k$th coherent state as a Gaussian distribution around the central frequency of the pulse $\omega_{P}^{0}$,
\begin{equation}
\alpha_{k}\pare{\omega_{k}} = \mathcal{K}e^{-\frac{\pare{\omega_{k} - \omega_{P}^{0}}^{2}}{2B_{P}^{2}}},
\tag{S25}
\end{equation}
with $B_{P}$ being the spectral bandwidth of the pulse and $\mathcal{K}$ a constant that can be related to the number of photons ($N_{P}$) in the pulse by means of the expression \cite{loudon_book}
\begin{equation}
N_{P}=\int\abs{\alpha_{k}\pare{\omega_{k}}}^{2}d\omega_{k}.
\tag{S26}
\end{equation}
Then, by substituting Eq. (S25) into (S26) we obtain the explicit form of $\mathcal{K}$, which allows us to write the complex amplitude of the $k$th coherent state as
\begin{equation}
\alpha_{k}\pare{\omega_{k}} = \pare{\frac{N_{P}}{\sqrt{\pi}B_{P}}}^{1/2}e^{-\frac{\pare{\omega_{k} - \omega_{P}^{0}}^{2}}{2B_{P}^{2}}}.
\tag{S27}
\end{equation}
Finally, by substituting Eq. (S27) into Eq. (\ref{gaussian-pulses}), the temporal shape of the pulse takes the form \cite{joel_book}
\begin{equation}
\varepsilon_{P}\pare{t-t_{P}} = \frac{\eta_{P}}{\sqrt{2\pi\sigma_P}}e^{-\frac{\pare{t-t_{P}}^{2}}{2\sigma_{P}^{2}}}e^{-i\omega_{P}^{0}\pare{t-t_{P}}},
\tag{S28}
\end{equation}
where the duration of the pulse is $\sigma_{P}=B_{P}^{-1}$, and its amplitude is given by
\begin{equation}
\eta_{P}=\pare{\frac{\hbar\omega_{P}^{0}N_{P}\sqrt{\pi}\sigma_{P}}{\epsilon_{0}cA}}^{1/2}.
\tag{S29}
\end{equation}
Notice that in order to obtain Eq. (S28), we have considered that the Gaussian bandwidth of the pulse is smaller than its central frequency, so the relation $\omega_{k} \simeq \omega_{P}^{0}$ can be assumed. This is a very good approximation, as $B_{P}$ is typically much smaller than the gap of the optical transitions which are resonant to the pulses. Note that, in the experimental proposal discussed in the main text, the central wavelength of the pump pulse is set to $775$ nm ($B_{P} \sim 44$ nm), which is indeed much larger than the gap between the molecular transitions $\Delta\lambda = 50$ nm.

\section*{\normalsize{III. Many-molecule pump-probe signals}}

In this section, we derive the expression for many-molecule pump-probe signals, which is the typical experimental observable, by considering the simplest scenario in which the absorbing medium consists of a mixture of $N_{mol}$ independent dimers (with $J\neq 0$), where the singly and doubly excited states of the $i$th dimer are represented by $\ket{\alpha_{i}}$, $\ket{\beta_{i}}$, and $\ket{f_{j}}$, respectively. In this situation, as stated in the main text, one needs to take into consideration the position of the dimer that is being excited by the optical fields. Therefore, the expressions (S1)--(S7) need to be modified in order to introduce the spatially dependent phase of the fields that interact with the $i$th molecule at the $\bm{r}_{i}$ position.

Let us start with the \emph{excited-state absorption} signal by writing
\begin{equation}
S_{ESA}= \left\langle\Psi_{+P+P'}\right.\ket{\Psi_{+P+P'}},
\tag{S30}
\label{ESA-many}
\end{equation}
with
\begin{equation}
\begin{split}
\ket{\Psi_{+P+P'}} = &\sum_{i}\sum_{j\neq i}e^{i\bm{k}_{P}\cdot\bm{r}_{i}}e^{i\bm{k}_{P'}\cdot\bm{r}_{j}}\ket{\Psi_{+P}^{\pare{\alpha_{i}}}}\ket{\Psi_{+P'}^{\pare{\alpha_{j}}}} + \sum_{i}\sum_{j\neq i}e^{i\bm{k}_{P}\cdot\bm{r}_{i}}e^{i\bm{k}_{P'}\cdot\bm{r}_{j}}\ket{\Psi_{+P}^{\pare{\beta_{i}}}}\ket{\Psi_{+P'}^{\pare{\beta_{j}}}} \\
& + \sum_{i}\sum_{j \neq i}e^{i\bm{k}_{P}\cdot\bm{r}_{i}}e^{i\bm{k}_{P'}\cdot\bm{r}_{j}}\ket{\Psi_{+P}^{\pare{\alpha_{i}}}}\ket{\Psi_{+P'}^{\pare{\beta_{j}}}} + \sum_{i}\sum_{j\neq i}e^{i\bm{k}_{P}\cdot\bm{r}_{i}}e^{i\bm{k}_{P'}\cdot\bm{r}_{j}}\ket{\Psi_{+P}^{\pare{\beta_{i}}}}\ket{\Psi_{+P'}^{\pare{\alpha_{j}}}} \\
& + \sum_{i}e^{i\bm{k}_{P}\cdot\bm{r}_{i}}e^{i\bm{k}_{P'}\cdot\bm{r}_{i}}\ket{\Psi_{+P+P'}^{\pare{f_{i}}}},
\end{split}
\tag{S31}
\label{ESAstate-many}
\end{equation}
where the two-photon excited state [last term in Eq. (\ref{ESAstate-many})] is given by Eq. (S5), and the one-photon excited state is defined as
\begin{equation}
\ket{\Psi_{+P}^{\pare{m}}} = \frac{i}{\hbar}\mu_{mg}\int_{-\infty}^{\infty} dt_{1}e^{-i\pare{\omega_{g}-\omega_{m}}t_{1}}\hat{E}_{P}^{\pare{+}}\pare{t_{1}}\ket{m}\ket{\varphi_{i}}.
\tag{S32}
\end{equation}
Notice that the first two terms of Eq. (\ref{ESAstate-many}) correspond to the states where the pump and probe fields excite the same transition in different dimers; whereas the other two describe the excitation of different transitions in two different dimers. The last term describes the situation in which the pump and probe fields excite the same dimer, that is, the excited-state absorption signal of a single molecule.

Then, by substituting Eq. (\ref{ESAstate-many}) into Eq. (\ref{ESA-many}), we obtain
\begin{equation}
\begin{split}
S_{ESA} =& \abs{\Omega_{\alpha g}^{P}}^{2}\abs{\Omega_{\alpha g}^{P'}}^{2}\cor{N_{mol}\pare{N_{mol}-2} + \abs{X}^{2}} \\
& + \abs{\Omega_{\beta g}^{P}}^{2}\abs{\Omega_{\beta g}^{P'}}^{2}\cor{N_{mol}\pare{N_{mol}-2} + \abs{X}^{2}} \\
& + \abs{\Omega_{\alpha g}^{P}}^{2}\abs{\Omega_{\beta g}^{P'}}^{2} N_{mol}\pare{N_{mol}-1} + \abs{\Omega_{\beta g}^{P}}^{2}\abs{\Omega_{\alpha g}^{P'}}^{2} N_{mol}\pare{N_{mol}-1} \\
& + \Omega_{g\alpha}^{\overline{P}}\Omega_{g\beta}^{\overline{P'}}\Omega_{\beta g}^{P}\Omega_{\alpha g}^{P'}e^{i\omega_{\alpha\beta}T} \pare{\abs{X}^{2} - N_{mol}} \\
& + \Omega_{g\beta}^{\overline{P}}\Omega_{g\alpha}^{\overline{P'}}\Omega_{\alpha g}^{P}\Omega_{\beta g}^{P'}e^{-i\omega_{\alpha\beta}T} \pare{\abs{X}^{2} - N_{mol}} \\
& + N_{mol} \sum_{m,n=\alpha,\beta} \Omega_{gn}^{\overline{P}}\Omega_{nf}^{\overline{P'}}\Omega_{fm}^{P'}\Omega_{mg}^{P}e^{-i\omega_{mn}T},
\end{split}
\tag{S33}
\label{finalESA-many}
\end{equation}
where
\begin{equation}
X = \sum_{j}e^{i\pare{\bm{k}_{P} - \bm{k}_{P'}}\cdot \bm{r}_{j}}.
\end{equation}
The parameters $\omega_{\alpha\beta}$ and $\Omega_{mg}^{j}$ (with $j=P,P'$, and $m=\alpha,\beta$) follow the same definitions as presented in the main text. Notice that this type of expressions contains interferences between pathways involving the same or different molecules.

Now, for the \emph{stimulated emission} signal, we have
\begin{equation}
S_{SE}= \left\langle\Psi_{+P-P'}\right.\ket{\Psi_{+P-P'}},
\tag{S34}
\label{SE-many}
\end{equation}
where
\begin{equation}
\ket{\Psi_{+P-P'}} = \sum_{j}e^{i\bm{k}_{P}\cdot\bm{r}_{j}}e^{-i\bm{k}_{P'}\cdot\bm{r}_{j}}\ket{\Psi_{+P-P'}^{\pare{\alpha_{j}}}} + \sum_{j}e^{i\bm{k}_{P}\cdot\bm{r}_{j}}e^{-i\bm{k}_{P'}\cdot\bm{r}_{j}}\ket{\Psi_{+P-P'}^{\pare{\beta_{j}}}},
\tag{S35}
\label{SEstate-many}
\end{equation}
with the second-order state is defined as
\begin{equation}
\ket{\Psi_{+P-P'}^{\pare{m}}} = -\frac{1}{\hbar^{2}}\mu_{gm}\mu_{mg}\int_{-\infty}^{\infty}dt_{2}\int_{-\infty}^{\infty}dt_{1}e^{-i\pare{\omega_{m}-\omega_{g}}t_{2}}e^{-i\pare{\omega_{g}-\omega_{m}}t_{1}} \hat{E}_{P'}^{\pare{-}}\pare{t_{2}}\hat{E}_{P}^{\pare{+}}\pare{t_{1}}\ket{g}\ket{\varphi_{i}}.
\tag{S36}
\end{equation}
Notice that Eq. (\ref{SEstate-many}) contains only two terms, this is because for the stimulated emission process to take place, both fields must interact with the same dimer.
Now, by substituting Eq. (\ref{SEstate-many}) into Eq. (\ref{SE-many}), we obtain
\begin{equation}
\begin{split}
S_{SE} = - & \abs{X}^{2}\left[ \abs{\Omega_{\alpha g}^{P}}^{2}\abs{\Omega_{\alpha g}^{P'}}^{2}-\abs{\Omega_{\beta g}^{P}}^{2}\abs{\Omega_{\beta g}^{P'}}^{2} - \Omega_{g\alpha}^{\overline{P}}\Omega_{g\beta}^{\overline{P'}}\Omega_{\beta g}^{P}\Omega_{\alpha g}^{P'}e^{i\omega_{\alpha\beta}T} - \Omega_{g\beta}^{\overline{P}}\Omega_{g\alpha}^{\overline{P'}}\Omega_{\alpha g}^{P}\Omega_{\beta g}^{P'}e^{-i\omega_{\alpha\beta}T}\right. \\
&\hspace{10mm} - \left. \abs{\Omega_{g\alpha}^{\text{vac},\overline{P'}}\Omega_{\alpha g}^{P} }^{2} + \abs{\Omega_{g\beta}^{\text{vac},\overline{P'}}\Omega_{\beta g}^{P} }^{2} + 2\Delta_{\alpha\beta}\mathcal{R}\pare{\Omega_{g\alpha}^{\text{vac},\overline{P'}}\Omega_{\alpha g}^{P}\Omega_{g\beta}^{\overline{P}}\Omega_{\beta g}^{\text{vac},P'} }\right].
\end{split}
\tag{S37}
\label{finalSE-many}
\end{equation}
Notice that the last term of Eq. (\ref{finalSE-many}) corresponds to the quantum vacuum contributions from the laser pulses, which do not appear in the case where they are considered as classical fields (see appendix C of Ref. \cite{joel_book}). Generalizing what was explained in Supplementary section IB (right after Eq. (S21)), this quantum term exhibits static double-slit interferences between emission pathways involving quasidegenerate transitions. In our case, this refers to $\llav{\ket{\alpha_{i}}\rightarrow\ket{g},\ket{\alpha_{j}}\rightarrow\ket{g}}$ or $\llav{\ket{\beta_{i}}\rightarrow\ket{g}, \ket{\beta_{j}}\rightarrow\ket{g}}$ for the same molecule ($i=j$) or different ones ($i\neq j$); clearly, these interferences also occur for $\llav{\ket{\alpha_{i}}\rightarrow\ket{g},\ket{\beta_{j}}\rightarrow\ket{g}}$ if $\Delta_{\alpha\beta}=1$. Now, recall that $\Omega_{ij}^{\text{vac},\overline{P'}}$ depends on $\eta_{P'}^{\text{vac}}=\cor{\hbar\omega_{P'}^{0}\Gamma/(\epsilon_{0}cA)}^{1/2}$. Here the appropriate decoherence lifetime $\Gamma$ is the one associated with the $\ket{\alpha_{i}}\bra{\alpha_{j}}$, $\ket{\beta_{i}}\bra{\beta_{j}}$, or $\ket{\alpha_{i}}\bra{\beta_{j}}$ coherences, respectively. If $i=j$, $\Gamma \sim 1$ ns for $\ket{\alpha_{i}}\bra{\alpha_{j}}$ or $\ket{\beta_{i}}\bra{\beta_{j}}$, which coincides with the excited state fluorescence lifetime of a bright chromophore. For all other cases, the decoherence is mediated by the distinct evolution of the baths associated with each excited state (in the same or different molecules), which is typically on the order of $\Gamma \sim 400$ fs (see Ref. \cite{lee2007}). To take a pessimistic bound, we just assume the latter for the calculation in the main text.

Finally, we derive the signal for the \emph{ground-state bleach} process by writing
\begin{equation}
S_{\text{GSB}} = 2\mathcal{R}\llav{\left\langle \Psi_{+P-P+P'}\right.\ket{\Psi_{+P'}}},
\tag{S38}
\label{GSB-many}
\end{equation}
with
\begin{equation}
\begin{split}
\ket{\Psi_{+P-P+P'}} = &N_{mol}\left[ \sum_{j}e^{i\bm{k}_{P'}\cdot\bm{r}_{j}}\ket{\Psi_{+P-P+P'}^{(\alpha \rightarrow \alpha_{j})}} + \sum_{j}e^{i\bm{k}_{P'}\cdot\bm{r}_{j}}\ket{\Psi_{+P-P+P'}^{(\alpha \rightarrow \beta_{j})}} \right. \\
& \hspace{8mm} + \left. \sum_{j}e^{i\bm{k}_{P'}\cdot\bm{r}_{j}}\ket{\Psi_{+P-P+P'}^{(\beta \rightarrow \alpha_{j})}} + \sum_{j}e^{i\bm{k}_{P'}\cdot\bm{r}_{j}}\ket{\Psi_{+P-P+P'}^{(\beta \rightarrow \beta_{j})}} \right],
\end{split}
\tag{S39}
\label{GSBstate-many1}
\end{equation}
and
\begin{equation}
\ket{\Psi_{+P'}} = \sum_{j}e^{i\bm{k}_{P'}\cdot\bm{r}_{j}}\ket{\Psi_{+P'}^{(\alpha_{j})}} + \sum_{j}e^{i\bm{k}_{P'}\cdot\bm{r}_{j}}\ket{\Psi_{+P'}^{(\beta_{j})}},
\tag{S40}
\label{GSBstate-many2}
\end{equation}
where the third-order state in Eq. (S24) is defined as
\begin{equation}
\begin{split}
\ket{\Psi_{+P-P+P'}^{(n \rightarrow m)}} = -\frac{i}{\hbar^{3}} \mu_{mg}\mu_{gn}\mu_{ng}&\int_{-\infty}^{\infty} dt_{3} \int_{-\infty}^{\infty}dt_{2}\int_{-\infty}^{t_{2}}dt_{1}e^{-i\pare{\omega_{g}-\omega_{m}}t_{3}}e^{-i\pare{\omega_{n}-\omega_{g}}t_{2}}e^{-i\pare{\omega_{g}-\omega_{n}}t_{1}} \\
& \times \hat{E}_{P'}^{\pare{+}}\pare{t_{3}} \hat{E}_{P}^{\pare{-}}\pare{t_{2}}\hat{E}_{P}^{\pare{+}}\pare{t_{1}} \ket{m}\ket{\varphi_{i}}.
\end{split}
\tag{S41}
\end{equation}
Here, the symbol $(n \rightarrow m)$ stands for the process in which the pump acts twice on the $n$ transition of the dimer first, leaving it in its ground state. Subsequently, the probe acts on the same or different $m$ transition, leaving it in its corresponding excited state.

By substituting Eqs. (\ref{GSBstate-many1})-(\ref{GSBstate-many2}) into Eq. (\ref{GSB-many}), we obtain
\begin{equation}
\begin{split}
S_{\text{GSB}} = &-\abs{\Omega_{\alpha g}^{P}}^{2}\abs{\Omega_{\alpha g}^{P'}}^{2}N_{mol}^{2} - \abs{\Omega_{\beta g}^{P}}^{2}\abs{\Omega_{\beta g}^{P'}}^{2}N_{mol}^{2} \\
& -\abs{\Omega_{\alpha g}^{P}}^{2}\abs{\Omega_{\beta g}^{P'}}^{2}N_{mol}^{2}-\abs{\Omega_{\beta g}^{P}}^{2}\abs{\Omega_{\alpha g}^{P'}}^{2}N_{mol}^{2},
\end{split}
\tag{S42}
\label{finalGSB-many}
\end{equation}
where the first two terms refer to processes with the same molecular transition, while the last two describe processes involving two different transitions.

Collecting the results in Eqs. (\ref{finalESA-many}), (\ref{finalSE-many}) and (\ref{finalGSB-many}), we find that
\begin{equation}
\begin{split}
S_{PP'} = & - N_{mol}\cor{\abs{\Omega_{\alpha g}^{P}}^{2}\abs{\Omega_{f\alpha}^{P'}}^{2} + \abs{\Omega_{\beta g}^{P}}^{2}\abs{\Omega_{f\beta}^{P'}}^{2} } \\
& - N_{mol}\cor{2\abs{\Omega_{\alpha g}^{P}}^{2}\abs{\Omega_{\alpha g}^{P'}}^{2} + 2\abs{\Omega_{\beta g}^{P}}^{2}\abs{\Omega_{\beta g}^{P'}}^{2} + \abs{\Omega_{\alpha g}^{P}}^{2}\abs{\Omega_{\beta g}^{P'}}^{2} + \abs{\Omega_{\beta g}^{P}}^{2}\abs{\Omega_{\alpha g}^{P'}}^{2} } \\
& + N_{mol}\cor{\Omega_{\alpha g}^{P}\Omega_{g\beta}^{\overline{P}}\Omega_{f\alpha}^{P'}\Omega_{\beta f}^{\overline{P'}}e^{-i\omega_{\alpha\beta}T} + \text{c.c.}} \\
& - N_{mol}\cor{\Omega_{\alpha g}^{P}\Omega_{g\beta}^{\overline{P}}\Omega_{g\alpha}^{\overline{P'}}\Omega_{\beta g}^{P'}e^{-i\omega_{\alpha\beta}T} + \text{c.c.}} \\
& - \abs{\sum_{j}e^{-i\pare{\bm{k}_{P} - \bm{k}_{P'}}\cdot \bm{r}_{j}}}^{2} \left[ \abs{\Omega_{g\alpha}^{\text{vac},\overline{P'}}\Omega_{\alpha g}^{P} }^{2} + \abs{\Omega_{g\beta}^{\text{vac},\overline{P'}}\Omega_{\beta g}^{P} }^{2} + 2\Delta_{\alpha\beta}\mathcal{R}\pare{\Omega_{g\alpha}^{\text{vac},\overline{P'}}\Omega_{\alpha g}^{P}\Omega_{g\beta}^{\overline{P}}\Omega_{\beta g}^{\text{vac},P'} }\right],
\end{split}
\tag{S43}
\end{equation}
which is the result presented in Eq. (10) of the main text. Notice that owing to cancelations between $S_{ESA}$ and $S_{SE}$, coherences between $\ket{\alpha_{i}}$ and $\ket{\beta_{j}}$ do not survive unless $i=j$, as expected.
\vspace{0.5cm}


\end{widetext}

\end{document}